# The impact of MR-guided attenuation correction (compared to CT-based AC) on the diagnosis of anosmia based on 99m-Tc Ethyl-Cysteinate-Dimer SPECT images

Faeze Gholamiankhah[1], Samaneh Mostafapour[1], Seid Kazem Razavi-Ratki[2], Ali Asghar Parach[1*] and Hossein Arabi[3]

[1]Department of Medical Physics, Shahid Sadoughi University of Medical Sciences, Yazd, Iran.

[2]Division of Nuclear Medicine and Molecular Imaging, Department of Medical Imaging, Geneva University Hospital, CH-1211 Geneva 4, Switzerland.

[3]Department of Radiology, Faculty of Medicine, Shahid Sadoughi University of Medical Sciences, Yazd, Iran.

[*]**Corresponding Author:** Ali Asghar Parach, Ph.D., Department of Medical Physics, Shahid Sadoughi University of Medical Sciences, Yazd, Iran.
E-mail: aliparach@ssu.ac.ir

Tel: +983516240691

**Running title:** MR-guided attenuation correction in brain SPECT imaging

**Type of article:** Original Research

**Abstract**

*Background*: 99m-Tc Ethyl-Cysteinate-Dimer SPECT and MR imaging play a significant role in diagnosing anosmia. In this study, two-tissue class and three-tissue class attenuation maps (2C-MR and 3C-MR) obtained from MR images were compared with CT-based attenuation correction (CTAC). Afterward, the presence of hypo-perfusion in brain lobes was evaluated in SPECT images.

*Materials and Methods*: The 2C-MRAC map was generated through segmentation of T1-W MR images into air and soft-tissue, while in the 3C-MRAC map, the cortical bone was also considered. For investigating MRAC approaches, the difference between activity concentration (ACC) values was estimated in 144 volumes of interest. Ten normal and fourteen anosmic patients were compared by calculating the average normalized count and standard uptake value ratio parameters in the brain lobes.

*Results*: The comparison between attenuation correction strategies represented that MRAC images resulted in underestimation of the ACC values which was more substantial in the cortical area rather than in central regions (maximum 9% vs. 6% for 2C-MR and maximum 5.5% vs. 3.5% for 3C-MR). Nevertheless, there was a strong correlation between the MRAC and CTAC methods with a correlation coefficient of 0.7 for both 2C-MR and 3C-MR. The statistical analysis between normal and affected groups indicated the hypo-perfusion in the cortex of Lh_frontal, Rh and Lh_temporal lobes with p-values < 0.05.

*Conclusions*: Using MRAC resulted in underestimation of activity concentration which was partly eliminated by considering the cortical bone in the 3C-MR attenuation map. Hypo-perfusion was perceived in Frontal and Temporal lobes in SPECT-MRAC images of the anosmic group.

**Keywords:** Anosmia, SPECT, quantitative imaging, MRI, CT

# 1. Introduction

Anosmia is defined as the lack of olfactory perception or defect of smelling sensation. Some studies have suggested that around 5-20% of individuals suffering from olfactory dysfunctions have had a history of head trauma [1]. Impaired olfactory sense has a significant impact on the quality and extent of life satisfaction. Also, since the patients will not be able to detect hazards, their safety would be seriously threatened [2]. Therefore, attempts for proper diagnosis of anosmia are crucial.

MRI anatomical imaging can give information about the olfactory system and the central nervous system pathology [3]. The use of functional imaging such as SPECT provides information about cerebral metabolism and perfusion [4]. Previous studies have shown reliable results in diagnosing anosmia by employing SPECT [5].

In addition to visual interpretation, a quantitative analysis of SPECT images is a valuable utility contributing to reliability and replicability in the report of images. Attenuation and scatter correction are two important factors for enhancing the quality of the image and increasing the reliability of quantification in emission tomography [6]. Although attenuation correction (AC) based on CT images (CTAC) or transmission scans (TXAC) is well-known and considered as ground-truth, these methods pose an additional dose to patients [7]. Attenuation correction using MR images (MRAC) as an alternative approach has extensively been used in PET imaging owing to the rapid growth of PET/MR scanners over the years [8-10]. Since, unlike CT images, MRI intensities represent proton density and relaxation time properties of the tissue, the mapping between MRI intensities and linear attenuation coefficient is challenging and three approaches have been developed in this regard [8, 11]: A) segmentation-based methods, which rely on the bulk classification of the major tissue types followed by the assignment of the predefined attenuation coefficient for each tissue class [12-14]. B) Atlas-based approaches which have shown promise in the context of synthetic CT estimation from MR images [15-17]. C) Deep learning-based methods in which synthetic CT estimation from MR images using convolutional neural networks has been a very active field, leading to promising results in the context of PET/MR AC [18-22].

In their study, Zaidi et al. indicated that segmentation-based MRAC in brain PET could be used as the second-best approach in the future. It also has the potential to be used in brain SPECT imaging [23]. However, the application of MRAC in brain SPECT has remained to unfold [24].

This study aimed to evaluate the MR-guided segmentation-based attenuation correction (AC) in the brain SPECT imaging versus the CT-based AC as the standard approach. Moreover, the impact of including the cortical bone in the MR-derived attenuation map as a separate tissue class in a three-class AC map (3C-MR) was evaluated against the commonly used uniform or two-class AC map (2C-MR). Besides, the presence of anosmia-induced hypo-perfusion within the different brain lobes in the SPECT images was quantified/assessed in terms of standard uptake value ratio (SUVR) and average normalized count (ANC) for the different SPECT AC approaches.

## 2. Materials and Methods

### 2.1. Patient population

The clinical dataset included 24 subjects (4 females and 20 males). Out of this number, 10 individuals were normal (mean age: 37.4±12.7) and 14 had anosmia (mean age: 33.5±11.4). The patients only with trauma-induced anosmia were included in this study. The patients who underwent neurosurgery or radiotherapy in the head region were excluded from this study. To diagnose the individuals with anosmia, both the report of the nuclear medicine specialist and the result of the University of Pennsylvania Smell Identification Test (UPSIT) were regarded [25].

This study was conducted under the permission granted by the Ethics Committee of Shahid Sadoughi University of Medical Sciences, Yazd, Iran, with the registered ethics code of IR.SSU.MEDICINE.REC.1396.110.

### 2.2. Data acquisition

For the SPECT scan, 20-25 mCi 99mTc-ECD (Ethyl-Cysteinate-Dimer) was injected intravenously to the subjects after 40 min of rest in the supine position with closed eyes in a room with low ambient light and noise. After 1-1.5 h resting under the same conditions, 120 projections were acquired over 360˚ by SPECT dual-head (Phillips ADAC, Forte, Netherlands) equipped with a low-energy high-resolution collimator in a relatively dark and quiet imaging room. Scatter correction was carried out using 20% photopeak window centered on 140 keV. The primary reconstruction was performed via STIR software [26] through the ordered subset expectation maximum [27, 28] -one step late algorithm. The axial reconstructed images had a matrix of 129 ×129 and a voxel size of $4.72 \times 4.72 \times 4.72$ mm$^3$.

Structural MR imaging was performed through SIEMENS Avanto 1.5 Tesla MRI scanner. The parameters of axial MR images were as follows: TR=400 ms, TE=8.7 ms, TI=0 ms, flip angle=90 degree, matrix= 256×256×170, and voxel size of $1 \times 1 \times 1$ mm$^3$.

CT data acquisition was performed on Alexion 16 Multi-Slice scanner (Toshiba) at the energy of 120 kVp in the supine position. X-ray data were collected in a matrix of 512 ×512 × 171 voxels with a resolution of $0.449 \times 0.449 \times 0.799$ mm$^3$.

## 2.2. Attenuation map generation

*Three-class attenuation map (3C-MR):* Prior to the MR-guided attenuation map generation, the CT and MR images were registered to the corresponding SPECT images through the landmarks alignment. The T1-W images were segmented using the method employed by Zaidi et al. [23] into three regions representing the skull, soft tissue, and air cavities. The attenuation coefficients were calculated as follows: the tissue composition and density ρ (g/cm$^3$) were extracted for each region using the ICRU 44 report [29]. Next, the mass attenuation coefficients were found by introducing the composition of the tissues into the XCOM photon cross-section library [30] and specifying the photon energy (140 keV). Through multiplying the mass attenuation coefficients by the densities, linear attenuation coefficients of 0.25, 0.15, and 0 were obtained for the skull, soft tissue, and air cavities, respectively.

*Two-class attenuation map (2C-MR):* In order to create a 2C-MR attenuation map, head contours were delineated in the T1 images and an attenuation coefficient of 0.15 (equivalent to the soft tissue) was assigned to all the pixels inside the head contour.

*CT-based attenuation map:* The voxel values in the CT images representing Hounsfield units were mapped to attenuation coefficients proportional to gamma photons at the energy of 140 keV by implementing a bilinear model [31].

The final reconstruction of the SPECT images was performed by incorporating the MR-derived and CT-based attenuation maps within the reconstruction algorithm. The parameters used in the OSEM algorithm were 8 subsets and 14 iterations [32]. Gaussian smoothing was performed with 0.6 standard deviations as post-reconstruction processing [33, 34].

## 2.3 Evaluation

In order to convert the voxel values to activity concentration, the calibration factor was calculated according to the technique used by Willowson et al. [35]. For this purpose, image acquisition was performed from a point source Tc-99m with the activity of 1 mCi in the air. Furthermore, regarding the fluctuations associated with the time interval between the injection and the starting time of image acquisition, all the data were decay corrected into the injected time.

For the evaluation of the 2C-MR and 3C-MR AC maps against the CT-based AC, 24 VOIs were defined on each image set including the cortex and the central parts of occipital, parietal, temporal, and frontal lobes as well as the cerebellum plus other areas such as cingulate and insula in each brain hemisphere. Then, the average count in each VOI was calculated to obtain activity concentration (ACC) using Eq. 1.

$$ACC\ (VOI) = mean\ count\ (VOI)/(CF \times v\ (VOI)) \qquad (1)$$

In this equation, *CF* and *v* indicate the calibration factor and volume of the VOI, respectively.
Assuming the SPECT-CTAC data as reference, the relative percentage error (*RE*) of ACC resulting from the SPECT-2C-MRAC and SPECT-3C-MRAC versus SPECT-CTAC values were calculated using Eq. 2 in the central and cortex regions of the brain for all VOIs:

$$\% \, RE \, (VOI) = (ACC_{(MR-AC)} - ACC_{(CT-AC)}) \times 100 / ACC_{(CT-AC)} \qquad (2)$$

Moreover, for evaluating the effect of 2C-MR and 3C-MR AC maps on the activity concentration of the reconstructed SPECT images, the relative percentage error ($RE_{MRAC}$) values of the SPECT-2C-MRAC were compared versus SPECT-3C-MRAC utilizing Eq. 3:

$$\% \, RE_{MRAC} \, (VOI) = (ACC_{(2C-MRAC)} - ACC_{(3C-MRAC)}) \times 100 / ACC_{(3C-MRAC)} \qquad (3)$$

Moreover, for investigating the correlation between the attenuation correction methods, a scatter plot was used in which the ACC data sets were plotted along the horizontal and vertical axes. The trendline obtained from each plot represents the correlation between the pair of considered methods.

For clinical assessment of the brain regions involved in anosmia, the presence of hypo-perfusion in the cortex of brain lobes was evaluated. Due to high inter-subject variability in the range of activity concentration data, of the use of this parameter did not reveal any differences between normal and anosmic groups. In recent years the Standard Uptake Value Ratio (SUVR) parameter has become a common parameter in conducting brain PET studies [36]. SUVR is defined as the SUV in the volume of interest divided by the SUV of the reference region [37]. In this study, based on the results of the research conducted by Eftekhari et al. [38], the cerebellum cortex was assumed as the reference, and the SUVR parameter was obtained as normalization of SUV (VOI) to SUV (cerebellum) for each subject.

Average normalized count (ANC) was another parameter whose validity for quantitative analysis and determining the hypo-perfusion in brain SPECT images was evaluated in this study. To obtain a normalized count, the authors considered a one-to-one correspondence in which the maximum pixel value in each image was re-scaled to 10, background to zero, and pixel values between 0-maximum were mapped to 0-10 using a linear function. To obtain ANC over a VOI, the average of pixel counts was calculated and used for normalization within the volume.

For investigating the functions of the left and right hemispheres of the brain, AUVR and ANC calculated in the left and right hemispheres in normal subjects were evaluated via an independent sample t-test. The analysis was performed for all three series of 2C-MR, 3C-MR, and CT attenuation corrected SPECT images. Also, using an independent sample t-test, statistical comparisons were made between the normal and abnormal subjects in terms of SUVR and ACC data obtained from VOIs across the cortex.

## 3. Results

Representative views of the CT and MR images together with 2C-MR and 3C-MR attenuation maps as well as corresponding SPECT-non AC, SPECT-CTAC, SPECT-2C-MRAC, and SPECT-3C-MRAC images for a single normal patient are shown in Fig. 1. The relative error percentages calculated based on ACC values in different VOIs, comparing the attenuation correction methods are reported in Figs. 2 and 3. In Fig. 2, CTAC and in Fig. 3, 3C-MRAC approaches are considered as the reference for calculating the relative error. According to plot 3A, the use of the 2C-MR AC map in comparison to CT-AC has resulted in underestimation of activity concentration in the central regions of the brain, which rose up to 6%. However, when approaching the cortical regions of the brain, as can be seen in part C, the extent of the underestimation increases up to 9%. According to the values of the plot 3B and 3D, when the 3C-MR AC map, containing bone tissue, is considered, the extent of underestimation diminishes both in the central and external regions of the brain, whereby the maximum values reach around 3.5% and 5.5%. Nevertheless, the extent of the underestimation is still lower in the central regions compared to the external regions. Furthermore, as it is indicated in plots A and B in Fig. 3, it can be inferred that the implementation of 2C-MR AC map in comparison with 3C-MR AC map in which the attenuation coefficient associated with cortical bone is not considered, leads to underestimation in the activity concentration. The amount of underestimation is more significant for VOIs located in the cortical area in comparison with VOIs located in central areas of the head (4.8% vs. 3%).

In the scatterplots in Fig. 4, linear regression plots have been drawn representing the correlation between the pair ACC values obtained from SPECT images (including both normal and affected patients) for the entire VOIs. In plots A and B, in which 2C-MRAC and 3C-MRAC are compared with CTAC, respectively, the trend lines have a slope of 0.98 and 0.96 with $R^2=0.72$ and 0.71. Also, the comparison of 2C-MRAC and 3C-MRAC approaches is demonstrated in plot C that has a slope of 1.02 and $R^2=0.99$. Thus, it can be concluded that despite the underestimation between the data resulting from SPECT-MRAC in comparison with SPECT-CTAC, there is a high correlation and a strong linear relationship between the corresponding values. The same strong correlation is obtained between the 2C-MRAC and 3C-MRAC values. Therefore, attenuation correction using the MR images, whether using 2 or 3 tissue classes including the bone tissue, will offer acceptable results. Furthermore, SUVR which is defined as the ratio of two SUV measurements and ANC parameters which is obtained by normalizing the counts, allowed for a reliable quantitative analysis of SPECT images.

Table 1 presents the p-values resulting from the independent sample t-test implemented to investigate the functions of the left and right hemispheres of the brain in normal subjects. According to the fact that all the estimated p-values are higher than 0.5, it can be inferred that in the absence of a disease impairing the normal pattern of brain perfusion, the activity of the left and right hemispheres in each lobe has no significant difference. This notion is true about all three series of 2C-MR, 3C-MR, and CTAC attenuation corrected SPECT images. Also, it is clear that the use of both SUVR and ANC metrics will yield the same statistical results. The mean coefficient of variation for ANC and SUVR was 0.07 and 0.6, respectively.

The results used for comparing brain perfusion between the normal and abnormal subjects in respective VOIs across the cortex are demonstrated in Tables 2 and 3. Table 2 reports the mean, standard deviation, and p-value related to each pair based on the SUVR parameter, and Table 3 reports these values based on the ANC parameter.

The data in Table 2 shows that the p-value in the cortex of Lh_frontal, Lh_temporal, and Rh_temporal is less than 0.05; thus, the SUVR parameter has a significant difference between the normal and anosmic groups. As such, it can be concluded that anosmia results in hypo-perfusion in these regions. It is also observed that in the Rh-frontal region, the p-value is close to the assumed threshold. The values in Table 3 also indicate similar interpretations. Therefore, both SUVR and ANC parameters enable distinguishing between the normal and abnormal groups as well as predicting the brain regions involved in the anosmia. Additionally, it is clear that the use of both data series, SPECT-2C-MRAC and SPECT-3C-MRAC, offers the same results in this assessment, confirming the accuracy of applying the 2C-MR attenuation correction.

## 4. Discussion

Early and accurate diagnosis of anosmia is essential for efficient patient management wherein the quantitative analysis of functional imaging data plays a significant role [4]. Attenuation correction is an essential prerequisite in quantitative SPECT imaging. In the absence of attenuation maps obtained from transmission scans (regarded as gold-standard), MR images have been employed to achieve the task of attenuation correction [8, 10, 14]. However, since MR signals do not directly provide electron density information, generating an attenuation map from MR is challenging. So far, various methods have been proposed/implemented to address this issue in SPECT and PET imaging including tissue segmentation, template-based, and machine learning-based approaches [8]. This study deals with the accuracy of segmentation-based MR-guided AC in SPECT imaging for the task of quantitative brain imaging assuming the CTAC as the standard method.

In a study, Anderson et al. [39] investigated water-only and water-only plus bone MRAC in comparison with CTAC in brain PET images. Based on their findings, neglecting the bone tissue in the AC map gave rise to underestimation of activity concentration up to 25% in the cortical regions and 5-10% in the soft-tissue regions. Inclusion of the bone tissue in the attenuation map remarkably reduced the quantitative bias and no systematic underestimation of the activity concentration was observed. In a similar study, Wagenknetch et al. [40] compared the segmentation-based MRAC with CTAC for brain PET imaging and reported systematic underestimation of activity concentration in the brain. They have also concluded that if the assigned AC coefficients are in good agreement with the patient-specific (or at least population-based) bone density, MRAC, and CTAC methods would lead to similar results. The comparison of MR-AC and CTAC approaches in our study exhibited the underestimation of the ACC values, particularly when 2C-MRAC approach (compared to 3C-MRAC) was employed (5.5% vs. 9%). The activity underestimation was more significant/severe in cortical regions rather than soft-tissue content regions.

Most MR-guided studies were performed in PET imaging while there is a limited number of publications dealing with SPECT imaging [14, 24, 41]. Deep learning-based AC approaches have also been proposed to perform AC in SPECT imaging. Nguyen et al. employed a 3-dimensional deep learning network that takes 3D patches of non-AC SPECT images as input of the network to directly estimate attenuation corrected SPECT images (performing AC in the image domain) [42]. Similarly, Mostafapour et al. investigated the application of the attenuation correction for MPI-SPECT imaging in the image space (not in projection space nor within image reconstruction) using a 2-dimensional neural network. They also implemented Chang's attenuation correction method (which is equivalent to the 2-class attenuation map) to provide a baseline for the assessment of the deep learning approach. The clinical evaluation, as well as quantitative results, demonstrated excellent performance of the deep learning approach (negligible bias) while systematic activity concentration bias was observed when Chang's method was used [43].

In a study conducted by Zaidi et al. [23], a segmentation-based MRAC was compared with transmission-based AC in brain PET scans. They reported a strong correlation between the two attenuation correction approaches and the reliability of the MR-AC approach for quantitative analysis in the PET images. They also concluded that since MR images bear low noise levels compared to transmission images, the resulting AC maps from MR images would induce less noise in the PET images. These findings are in agreement with the observations made in this study as the evaluated MRAC approaches exhibited a high correlation with CTAC.

Eftekhari et al. [38] in their investigation of 99mTc-ECD brain SPECT scans indicated that there is no remarkable difference in the normal pattern of cerebral activation regarding the hemispheric activity. Likewise, the assessment of perfusion in normal subjects in this paper revealed no significant difference in the corresponding VOIs in the left and right hemispheres.

Gerami et al. [44] investigated the brain perfusion state of anosmic patients and normal controls through 99mTc-ECD SPECT scanning. They supposed the Orbitofrontal lobe as VOI and demonstrated the fact that anosmia following a head trauma led to hypo-perfusion in this region. In another analogous experiment on patients with posttraumatic impaired smell, Eftekhari et al. [38] reported hypo-perfusion in the Orbital

Frontal Cortex, Inferior and Superior frontal Pole, Posterior Superior Frontal Lobe, and Parasagittal Area. Their research revealed no statistically significant difference in the pattern of cerebral activation in terms of patients' gender. A semi-quantitative analysis performed by Atighechi et al. [45] observed hypo-perfusion in the Frontal, left Temporal and left Parietal in anosmic patients. The statistical analysis in our study indicated that anosmia leads to hypo-perfusion in the Rh-frontal and temporal regions. Moreover, the obtained p-value for Lh_frontal is very close to the assumed threshold. Clearly, the results obtained from the 2C-MRAC and 3C-MRAC methods are similar to each other.

Although attenuation correction of the brain SPECT images using the AC maps generated from MR images led to acceptable results in comparison with the CT-based attenuation correction, this approach faces a number of limitations and challenges. The cortical bone which plays a significant role in attenuation of gamma photons has a very weak or no signal in the T1-W MR images. As such, the cortical bone could not be properly detected in the T1-W MR images, which complicates the process of including a separate class for bony structures into the resulting AC maps. To address this issue, an ultrashort echo time (UTE) pulse sequence can be used in MR imaging which can provide strong signals for the bony structures. The other solution would be the use of a template or atlas-based methods to include the bone tissue in the MR-derive AC maps [39, 40]. However, these approaches exhibit sub-optimal performance for the subjects with abnormal anatomies. The other limitation of this work is associated with the MR and SPECT images acquired non-concurrently which may cause misalignment errors. The different patient positionings within the MR and SPECT scans would challenge the perfect alignment of these two images. Nonetheless, it should be noted that this effect is less significant for brain imaging because of the rigidity of the head structures. It is recommended that further studies involve more subjects in order to make possible determining the threshold for distinguishing anosmic patients as well as the severity of the disease.

## 5. Conclusion

Both SPECT and MR imaging play a significant role in diagnosing anosmia. Using MRAC methods in comparison with CTAC in brain SPECT led to an underestimation in activity concentration which was more considerable in the cortex area and especially when ignoring cortical bone. Nevertheless, the two methods revealed a strong correlation with each other. Furthermore, the excellent soft-tissue contrast provided by MR imaging allowed defining more precise VOIs on brain scans to achieve more accuracy in the quantitative assessment of brain SPECT. The results of using the SUVR parameter for evaluation of brain regions involved in anosmia illustrated hypo-perfusion in the cortex of Frontal and Temporal lobes. Using data derived from ANC yielded similar results.

**Acknowledgments:** Hereby the contributions of Jelodari and all the personnel of the nuclear medicine department of Shahid Sadoughi Hospital in Yazd in collecting the necessary data are appreciated.

**Compliance with Ethical Standards:**
**Funding sources:** This research did not receive any specific grant funding from agencies in the public, commercial, or not-for-profit sectors.
**Conflict of interest:** The authors declare that there is no conflict of interest.
**Ethical approval:** The analysis of patients' data was performed with the permission of the Ethics Committee of our university.
**Informed consent:** Informed consent was obtained from all individual participants included in the study to investigate their images for this project.

**Table 1.** Comparing the functions of the left and right brain hemispheres in normal subjects using SUVR and ANC parameters

|  | 2C-MRAC | | 3C-MRAC | | CTAC | |
|---|---|---|---|---|---|---|
|  | SUVR | ANC | SUVR | ANC | SUVR | ANC |
| Frontal_ctx | 0.97 | 0.92 | 0.76 | 0.79 | 0.67 | 0.33 |
| Occipital_ctx | 0.55 | 0.65 | 0.54 | 0.64 | 0.66 | 0.67 |
| Temporal_ctx | 0.73 | 0.79 | 0.68 | 0.73 | 0.94 | 0.82 |
| Parietal_ctx | 0.39 | 0.44 | 0.50 | 0.53 | 0.70 | 0.66 |

SUVR standard uptake value ratio, ANC average normalized count, 2C-MRAC two-tissue class MRI attenuation correction, 3C-MRAC three-tissue class MRI attenuation correction, ctx cortex

**Table 2.** Comparison of the standard uptake value ratio in the brain lobes between the normal and abnormal subjects.

|                  | 2C-MRAC         |                 |         | 3C-MRAC         |                 |         |
|------------------|-----------------|-----------------|---------|-----------------|-----------------|---------|
|                  | Normal          | Abnormal        | p-value | Normal          | Abnormal        | p-value |
| Lh_frontal_ctx   | 0.88 ± 0.03     | 0.78 ± 0.04     | 0.03    | 0.91 ± 0.02     | 0.83 ± 0.09     | 0.008   |
| Rh_frontal_ctx   | 0.89 ± 0.04     | 0.81 ± 0.16     | 0.10    | 0.91 ± 0.04     | 0.85 ± 0.11     | 0.09    |
| Lh_occipital_ctx | 0.81 ± 0.04     | 0.78 ± 0.14     | 0.50    | 0.83 ± 0.04     | 0.82 ± 0.08     | 0.82    |
| Rh_occipital_ctx | 0.79 ± 0.09     | 0.73 ± 0.13     | 0.18    | 0.81 ± 0.08     | 0.77 ± 0.08     | 0.26    |
| Lh_temporal_ctx  | 0.86 ± 0.02     | 0.77 ± 0.12     | 0.01    | 0.87 ± 0.04     | 0.81 ± 0.05     | 0.01    |
| Rh_temporal_ctx  | 0.87 ± 0.06     | 0.77 ± 0.12     | 0.03    | 0.88 ± 0.06     | 0.82 ± 0.05     | 0.04    |
| Lh_parietal_ctx  | 0.86 ± 0.07     | 0.79 ± 0.17     | 0.20    | 0.89 ± 0.06     | 0.83 ± 0.12     | 0.10    |
| Rh_parietal_ctx  | 0.84 ± 0.06     | 0.78 ± 0.16     | 0.23    | 0.86 ± 0.05     | 0.82 ± 0.12     | 0.33    |

2C-MRAC two-tissue class MRI attenuation correction, 3C-MRAC three-tissue class MRI attenuation correction, Lh left hemisphere, Rh right hemisphere, ctx cortex

**Table 3.** Comparison of the average normalized count in the brain lobes between the normal and abnormal subjects

|  | 2C-MRAC | | | 3C-MRAC | | |
|---|---|---|---|---|---|---|
|  | Normal | Abnormal | p-value | Normal | Abnormal | p-value |
| Lh_frontal_ctx | 6.20 ± 0.32 | 5.57 ± 0.90 | 0.02 | 6.34 ± 0.32 | 5.71 ± 0.83 | 0.03 |
| Rh_frontal_ctx | 6.23 ± 0.32 | 5.77 ± 0.90 | 0.09 | 6.33 ± 0.28 | 5.81 ± 0.91 | 0.09 |
| Lh_occipital_ctx | 5.70 ± 0.34 | 5.54 ± 0.47 | 0.37 | 5.76 ± 0.31 | 5.63 ± 0.47 | 0.47 |
| Rh_occipital_ctx | 5.58 ± 0.75 | 5.18 ± 0.53 | 0.14 | 5.64 ± 0.75 | 5.28 ± 0.54 | 0.18 |
| Lh_temporal_ctx | 6.03 ± 0.33 | 5.49 ± 0.45 | 0.005 | 6.10 ± 0.35 | 5.58 ± 0.48 | 0.009 |
| Rh_temporal_ctx | 6.09 ± 0.46 | 5.48 ± 0.40 | 0.002 | 6.15 ± 0.50 | 5.62 ± 0.35 | 0.006 |
| Lh_parietal_ctx | 6.03 ± 0.49 | 5.60 ± 0.90 | 0.19 | 6.13 ± 0.40 | 5.69 ± 0.90 | 0.15 |
| Rh_parietal_ctx | 5.91 ± 0.52 | 5.57 ± 0.87 | 0.28 | 5.98 ± 0.49 | 5.65 ± 0.87 | 0.29 |

2C-MRAC two-tissue class MRI attenuation correction, 3C-MRAC three-tissue class MRI attenuation correction, Lh left hemisphere, Rh right hemisphere, ctx cortex

**Figures captions**

**Fig. 1.** Attenuation map generation: A) T1_W MRI used for generating MRAC maps. B) 2C segmented MRI. C) 3C segmented MRI. D) CT. Attenuation correction maps: E) 2C-MRAC map. F) 3C-MRAC map. G) CTAC map. H) Primary reconstructed SPECT without AC. I) SPECT-2C-MRAC. J) SPECT-3C-MRAC. K) SPECT-CTAC.

**Fig. 2.** The extent of relative error percentage: A) 2C-MRAC vs. CTAC in the central cerebellar region. B) 3C-MRAC vs. CTAC in the central cerebellar region. C) 2C-MRAC vs. CTAC in the cortical region. D) 3C-MRAC vs. CTAC in the cortical region.

**Fig. 3.** The extent of relative error percentage in 2C-MRAC vs. 3C-MRAC as reference A) in the central cerebellar region, and B) in the cortical region.

**Fig. 4.** Evaluation of correlation between attenuation correction methods through the scatter diagram. A) SPECT-CTAC and SPECT-2C-MRAC. B) SPECT-CTAC and SPECT-3C-MRAC. C) SPECT-2C-MRAC and SPECT-3C-MRAC.

**Fig. 1**

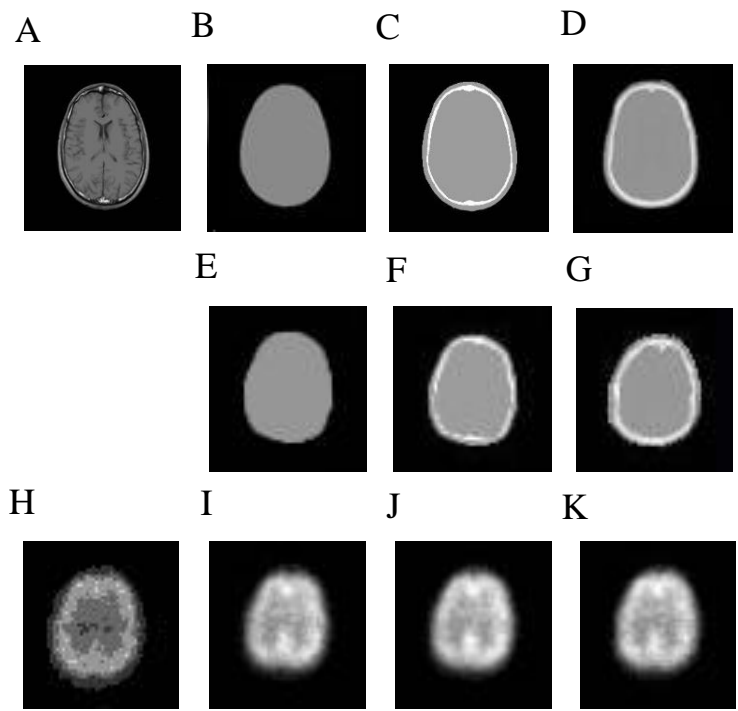

**Fig. 2**

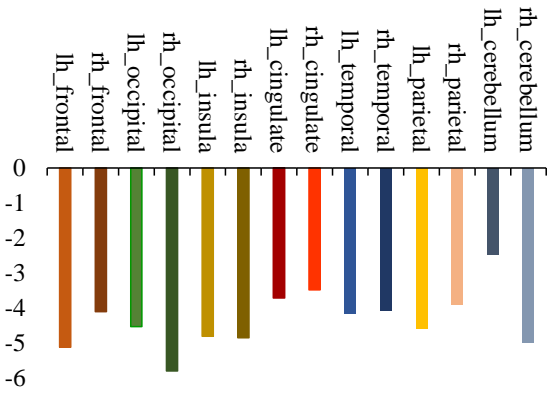
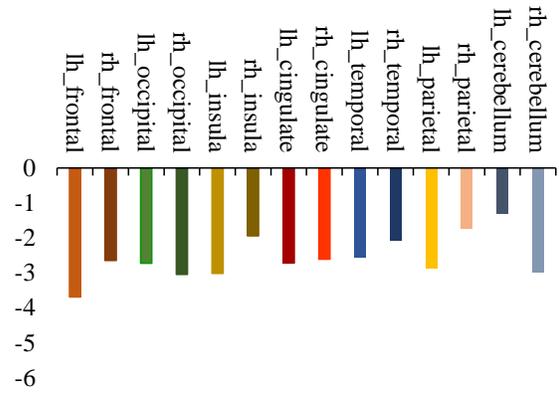
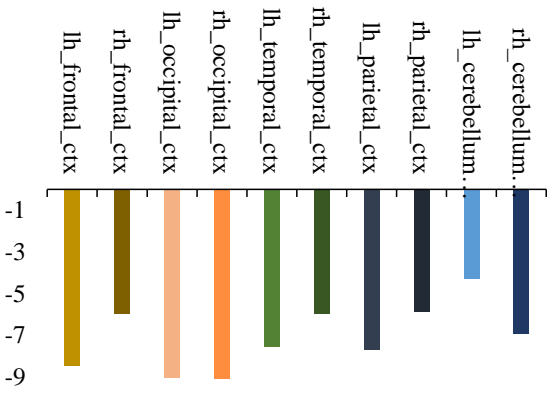
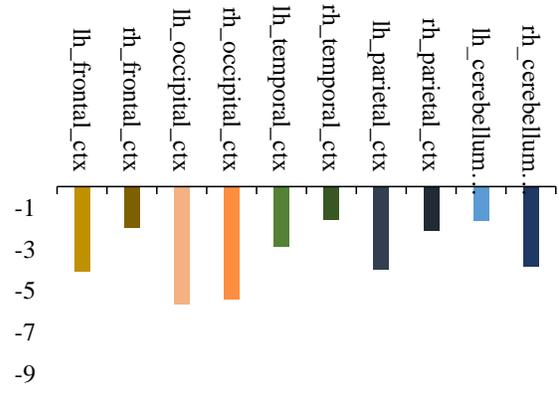

**Fig. 3**

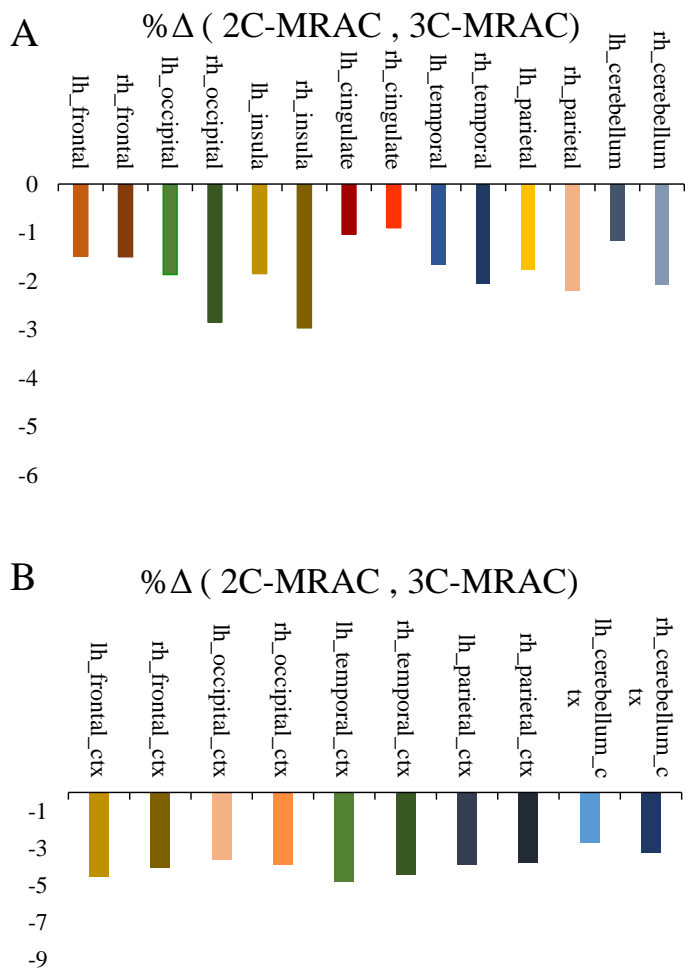

**Fig. 4**

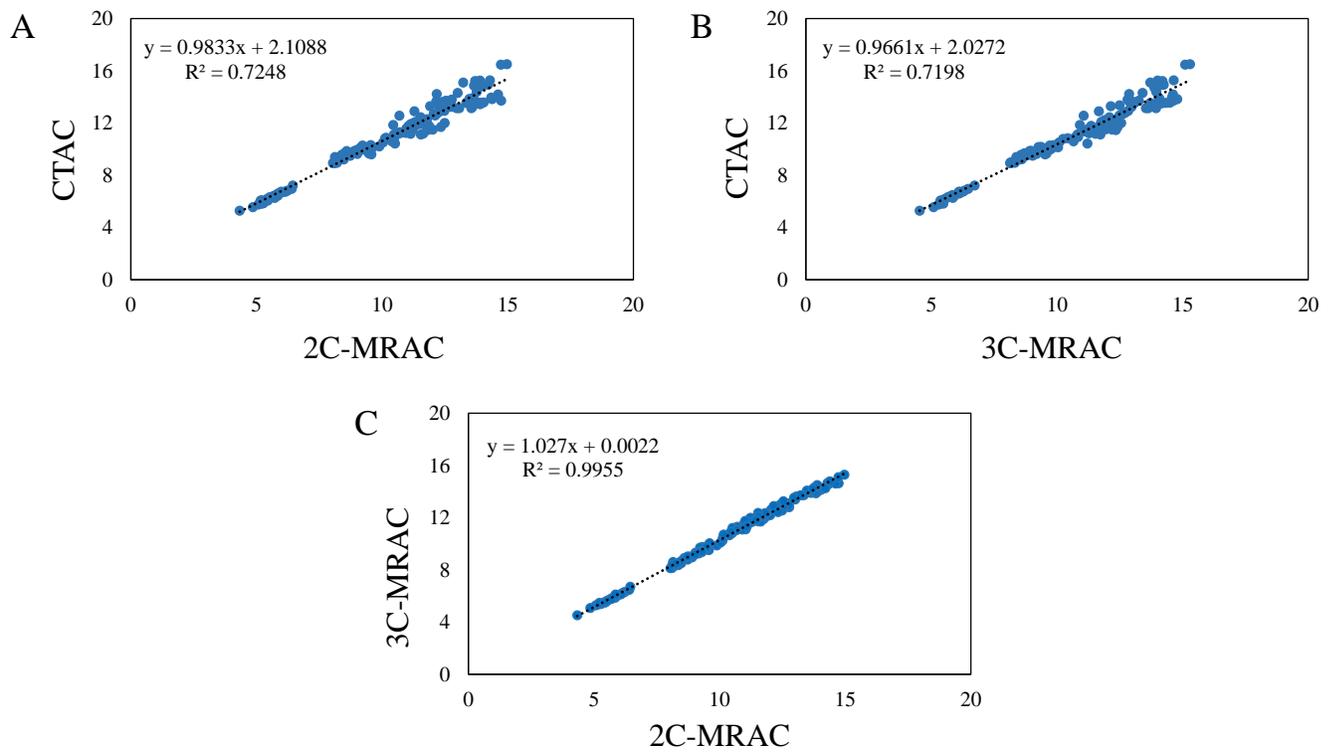